\documentclass[11pt,a4paper]{article}
\usepackage[utf8]{inputenc}
\usepackage{amsmath,amssymb,amsfonts,mathrsfs}
\usepackage{geometry}
\usepackage{hyperref}
\usepackage{microtype}

\usepackage{calrsfs}
\DeclareMathAlphabet{\pazocal}{OMS}{zplm}{m}{n}

\def\bea{\begin{eqnarray}}
\def\eea{\end{eqnarray}}
\def\ben{\begin{eqnarray*}}
\def\een{\end{eqnarray*}}
\def\t{\widetilde}
\newcommand{\E}{\mathcal{E}}

\def\L{\mathcal{L}}

\def\N{\mathcal{N}}

\def\I{\mathtt{I}}

\def\eps{{\cal{\varepsilon}}}

\begin{document}

\pagestyle{empty}
\vskip-10pt
\vskip-10pt
\hfill %{\tt hep-th/yymmnnn}
\begin{center}
\vskip 3truecm
{\Large \bf
In light of the Henneaux-Teitelboim action
}
\vskip 2truecm
{\large \bf
Andreas Gustavsson}
\vspace{1cm} 
\begin{center} 
Physics Department, University of Seoul, Seoul 02504 KOREA
\end{center}
\end{center}
\vskip 2truecm
{\abstract We construct an action similar to the Henneaux-Teitelboim action on a six dimensional Lorentzian manifold where the time direction is chosen to be lightlike and generated by a conformal lightlike Killing vector $U$. We show that such an action exists only if there is another commuting light-like conformal Killing vector $V$ with properties that can be realized by the Dirac currents of two conformal Killing spinors. We extend the action to the $(1,0)$ supersymmetric tensor multiplet.}

\vfill
\vskip4pt
\eject
\pagestyle{plain}

\section{Introduction}
In this paper we will construct an action for the abelian selfdual tensor field $H = d B$ in six dimensions where two lighlike direction are separated from the remaining four transverse directions. We are thus asking for an action that upon variation of $B$ gives us the selfduality equations of motion
\bea
L &=& 0
\eea
where we define 
\bea
L &=& H - * H
\eea
as an off-shell anti-selfdual tensor field. The story begins in 1988 when Henneaux and Teitelboim (HT) presented an action for a selfdual tensor field in which the time direction is selected as a special direction \cite{Henneaux:1988gg}. They constructed an action that gives the  selfdual equations of motion by utilizing a gauge symmetry of the action. The action is not manifestly Lorentz invariant in flat Minkowski spacetime. It arises naturally in a Hamiltonian framework with a timelike Killing vector field. Later research has been made on revealing either a hidden Lorentz symmetry \cite{Perry:1996mk}, \cite{Chen:2010jgb}, \cite{Huang:2011np} or an action with manifest Lorentz symmetry \cite{Pasti:1996vs}, \cite{Sorokin:2025ezs}. It has been covariantized in a two-metric formulation by Hull \cite{Hull:2023dgp} by extending the work by Sen \cite{Sen:2019qit}. Although these recent developments based on Sen's approach seem distinct from the PST formulation, it was shown in \cite{Vanichchapongjaroen:2024tkj} that two PST actions can be mapped to Hull's action. By the work of \cite{Sorokin:2025ezs} one may obtain the PST action from the Hamilonian of a non-selfdual tensor field that leads to the HT action and the PST action. We can now understand that all these different formulations can be traced back to the HT action in one way or another. In the original HT action the time direction was assumed to be timelike. We will examine whether we can construct an analog of the HT action when the time direction is taken to be along a lightlike direction. We follow the spirit of original formulation of Henneaux and Teitelboim where we will accept an explicit breaking of the Lorentz group in flat Minkowski spacetime of the action and proceed to obtain a curved spacetime generalization. One may now want to apply the steps in  \cite{Sorokin:2025ezs} and derive a PST action with a lightlike time direction instead of a timelike time direction. However, due to certain geometric conditions that we obtain here, we do not expect this approach to be completely straightforward \cite{Ko:2015zsy}. 

\section{Two lightlike vector fields}
We would like to obtain an action for a selfdual three-form tensor field in six-dimensions that is of a similar type as the action that was obtained by Henneaux and Teitelboim \cite{Henneaux:1988gg} on a generic Lorentzian six-manifold, but where we instead of taking time along a timelike direction, will let time be along a lightlike direction. 

We introduce a lightlike vector field $U$ that in flat Minkowski spacetime with metric $ds^2 = 2 dx^+ dx^- + dx^i dx^i$ for $i = 1,...,4$ will be $U = \partial_+$ and the time coordinate will be $x^+$. In order to not encounter a singular Lorentzian metric we also need to introduce another lightlike vector field $V$ that in flat Minkowski spacetime would be $V = \partial_-$. 

We will start without making any assumptions regarding these vector fields other than that they are both lightlike and that they have an everywhere nonvanishing inner product,
\bea
\N = g_{MN} U^M V^N > 0
\eea
as necessary for the metric to be everywhere non-degenerate with a non-vanishing determinant. Here $\N$ need not be constant. We do not assume from the outset that any of these vector fields are conformal Killing vector fields. For notational convenience, we define
\bea
\Omega_V &:=& \nabla_M V^M\cr
\Omega_U &:=& \nabla_M U^M
\eea
Given these two lightlike vector fields, the metric tensor may be expanded as
\bea
g_{MN} &=& \frac{1}{\N} (U_M V_N + V_M U_N) + \t{g}_{MN}
\eea
By contracting both sides with $V^N$ and using that $V$ is lightlike it follows that $\t{g}_{MN} V^N = 0$ and similarly by contracting with $U^N$ and using that $U$ is lightlike, we get $\t{g}_{MN} U^N = 0$. The inverse metric can likewise be expanded as
\bea
g^{MN} &=& \frac{1}{\N} (U^M V^N + V^M U^N) + \t{g}^{MN}
\eea
and again one can show that $\t{g}^{MN} V_N = 0$ and $\t{g}^{MN} U_N = 0$. We also can show that 
\bea
\t{g}^{MN} \t{g}_{NP} &=& P^M{}_P
\eea
where 
\bea
P^M{}_P &=& \delta^M_P - Q^M_P\cr
Q^M_P &=& \frac{1}{\N} (U^M V_P + V^M U_P)
\eea
Here $P$ projects onto the four-dimensional transverse space orthogonal to the two lightlike vectors $U$ and $V$ and $Q$ projects onto the space spanned by $U$ and $V$. Nothing in this exposition so far requires these vector fields to be conformal Killing vector fields. 

We may introduce one-form fields 
\bea
u &=& \frac{U_M}{\N} dx^M\cr
v &=& \frac{V_M}{\N} dx^M
\eea
and a frame
\bea
e^u &=& \sqrt{\N} u\cr
e^v &=& \sqrt{\N} v\cr
E^i &=& E^i_M dx^M
\eea
where $E^i_M E^i_N = \t{g}_{MN}$ so that $V^M E_M^i = 0$ and $U^M E_M^i = 0$. The metric expressed in this frame reads
\bea
ds^2 &=& 2 e^u e^v + E^i E^i
\eea
We define the volume form as 
\bea
*1 &=& e^v \wedge e^u \wedge E^1 \wedge E^2 \wedge E^3 \wedge E^4\cr
&=& \N v \wedge u \wedge *_4 1
\eea
where we the volume form on the transverse four-dimensional space is
\bea
*_4 1 &=& E^1 \wedge E^2 \wedge E^3 \wedge E^4
\eea
In Minkowski spacetime with $u = dx^-$ and $v = dx^+$ the sign convention here corresponds to $\eps^{+-} = 1$.

\section{The action}
We will make the following ansatz for the action
\bea
S &=& \int d^6 x \L
\eea
where we take the Lagrangian density to be a sum of three terms
\bea
\L &=& - \L_{\I} - \L_{\I\I} - \L_{\I\I\I}
\eea
where
\bea
\L_{\I} &=& \frac{\sqrt{-g}}{24} H_{MNP}^2 \cr
\L_{\I\I} &=& - \frac{\sqrt{-g}}{8\N} F^{MN} G_{MN}\cr
\L_{\I\I\I} &=& - \frac{\sqrt{-g}}{96} \eps^{MNPRST}  \left(F_{MN} \frac{U_P}{\N} + G_{MN} \frac{V_P}{\N}\right) H_{RST}
\eea
Here we define
\bea
F_{MN} &=& H_{MNP} V^P\cr
G_{MN} &=& H_{MNP} U^P
\eea
We require the action to be Weyl invariant. The action above is constructed so that each term is Weyl invariant by appropriate insertions of the normalization function $\N$ whose Weyl transformation is the same as that of the metric tensor,
\bea
g_{MN} &\rightarrow & e^{2\sigma} g_{MN}\cr
\N &\rightarrow & e^{2\sigma} \N
\eea
We will now obtain the equations of motion from this action by varying the gauge potential. We consider each term in the Lagrangian separately and drop all boundary terms as they appear upon integrations by parts.

\subsection{The variation of $\L_{\I}$} 
\bea
\delta \L_{\I} &=& \frac{\sqrt{-g}}{4} H^{MNP} \partial_M \delta B_{NP}\cr
&=& - \frac{\sqrt{-g}}{4} \left(\nabla_M H^{MNP}\right) \delta B_{NP}
\eea

\subsection{The variation of $\L_{\I\I}$}
\bea
\delta \L_{\I\I} &=& - \frac{\sqrt{-g}}{8\N} \left(G^{MN} \delta F_{MN} + F^{MN} \delta G_{MN}\right)
\eea
Here we note that
\bea
\delta F_{MN} &=& \L_V B_{MN} + 2 \partial_M (\delta B_{NP} V^P)\cr
\delta G_{MN} &=& \L_U B_{MN} + 2 \partial_M (\delta B_{NP} U^P)
\eea
so then 
\bea
\delta \L_{\I\I} &=& \frac{\sqrt{-g}}{8} \left(\Omega_V \frac{G^{MN}}{\N} + \L_V \left(\frac{G^{MN}}{\N}\right) + 2 \nabla_P \left(\frac{G^{PM}}{\N}\right) V^N\right) \delta B_{MN}\cr
&& + \frac{\sqrt{-g}}{8} \left(\Omega_U \frac{F^{MN}}{\N} + \L_U \left(\frac{F^{MN}}{\N}\right) + 2 \nabla_P \left(\frac{F^{PM}}{\N}\right) U^N\right) \delta B_{MN}
\eea
We now expand
\bea
\nabla_P \left(\frac{G^{PM}}{\N}\right) &=& \nabla_P H^{PMQ} \frac{U_Q}{\N} + \frac{1}{2} H^{PMQ} U_{PQ}\cr
\nabla_P \left(\frac{F^{PM}}{\N}\right) &=& \nabla_P H^{PMQ} \frac{V_Q}{\N} + \frac{1}{2} H^{PMQ} V_{PQ}
\eea
where we define
\bea
U_{MN} &=& \partial_M \left(\frac{U_N}{\N}\right) - \partial_N \left(\frac{U_M}{\N}\right)\cr
V_{MN} &=& \partial_M \left(\frac{V_N}{\N}\right) - \partial_N \left(\frac{V_M}{\N}\right)
\eea
Inserting these expansions, we get
\ben
\delta \L_{\I\I} &=& \frac{\sqrt{-g}}{8} \left(\Omega_V \frac{G^{MN}}{\N} + \L_V \left(\frac{G^{MN}}{\N}\right) + 2 \nabla_P H^{PMQ} \frac{U_Q V^N}{\N} - H^{MPQ} U_{PQ} V^N\right) \delta B_{MN}\cr
&& + \frac{\sqrt{-g}}{8} \left(\Omega_U \frac{F^{MN}}{\N} + \L_U \left(\frac{F^{MN}}{\N}\right) + 2 \nabla_P H^{PMQ} \frac{V_Q U^N}{\N} -  H^{MPQ} V_{PQ} U^N\right) \delta B_{MN}
\een
Let us now notice the following identities, which we prove in appendix \ref{ide},
\bea
\left(H^{MPQ} V^N - H^{NPQ} V^M\right) U_{PQ} &=& \frac{1}{2} F_{RS} U_{PQ} \eps^{RSPQMN}\cr
\left(H^{MPQ} U^N - H^{NPQ} U^M\right) V_{PQ} &=& \frac{1}{2} G_{RS} V_{PQ} \eps^{RSPQMN}
\eea
Inserting these identities we get
\bea
\delta \L_{\I\I} &=& \sqrt{-g} \Bigg[\frac{1}{8} \L_V \left(\frac{G^{RS}}{\N}\right) + \frac{1}{8} \Omega_V \frac{G^{RS}}{\N} \cr
&& - \frac{1}{32} \eps^{MNPQRS} F_{MN} U_{PQ}\cr
&& + \frac{1}{4} \frac{U_Q V^S}{\N} \nabla_P H^{PRQ} \Bigg]\delta B_{RS}\cr
&& + \sqrt{-g} \Bigg[\frac{1}{8} \L_U \left(\frac{F^{RS}}{\N}\right) + \frac{1}{8} \Omega_U \frac{F^{RS}}{\N} \cr
&& - \frac{1}{32} \eps^{MNPQRS} G_{MN} V_{PQ}\cr
&& + \frac{1}{4} \frac{V_Q U^S}{\N} \nabla_P H^{PRQ} \Bigg]\delta B_{RS}
\eea

\subsection{The variation of $\L_{\I\I\I}$}
Here we will obtain the variation of 
\bea
\L_{\I\I\I} &=& \frac{\sqrt{-g}}{96} \eps^{MNPRST} H_{RST} \left(F_{MN} \frac{U_P}{\N} + G_{MN} \frac{V_P}{\N}\right)
\eea
When we vary $B_{MN}$ inside $\L_{\I\I\I}$ we notice that the combination $\sqrt{-g} \eps^{MNPQRS}$ is just a numerical, either $+1$ or $-1$, so no derivative acts on this combination nontrivially. This means that we can perform integration by parts easily and get
\bea
\delta \L_{\I\I\I} &=& - \frac{\sqrt{-g}}{32} \eps^{MNPRST} \delta B_{ST} \partial_R \left(F_{MN} \frac{U_P}{\N} + G_{MN} \frac{V_P}{\N}\right)\cr
&& - \L_V \left(\frac{\sqrt{-g}}{96} \eps^{MNPRST} H_{RST} \frac{U_P}{\N}\right) \delta B_{MN}  \cr
&& - \L_U \left(\frac{\sqrt{-g}}{96} \eps^{MNPRST} H_{RST} \frac{V_P}{\N}\right) \delta B_{MN}  \cr
&& - \frac{\sqrt{-g}}{96} \eps^{MNPRST} H_{RST} \delta B_{NQ} V^Q U_{MP}\cr
&& - \frac{\sqrt{-g}}{96} \eps^{MNPRST} H_{RST} \delta B_{NQ} U^Q V_{MP}
\eea
We now note the following modified Bianchi identities,
\bea
3\partial_{[R} F_{MN]} &=& \L_V H_{MNR}\cr
3\partial_{[R} F_{MN]} &=& \L_U H_{MNR}
\eea
and then we get 
\bea
\delta \L_{\I\I\I} &=& - \L_V \left(\frac{\sqrt{-g}}{48} \eps^{MNPRST}   H_{MNR} \frac{U_P}{\N}\right) \delta B_{ST}\cr
&& - \L_U \left(\frac{\sqrt{-g}}{48} \eps^{MNPRST} H_{MNR} \frac{V_P}{\N}\right) \delta B_{ST}\cr
&& - \frac{\sqrt{-g}}{64} \eps^{MNPRST} \delta B_{ST} \left(F_{MN} U_{RP} + G_{MN} V_{RP}\right)\cr
&& - \frac{\sqrt{-g}}{96} \eps^{MNPRST} H_{RST} \delta B_{NQ} V^Q U_{MP}\cr
&& - \frac{\sqrt{-g}}{96} \eps^{MNPRST} H_{RST} \delta B_{NQ} U^Q V_{MP}\cr
&& + \frac{\sqrt{-g}}{96} \eps^{MNPRST}   H_{MNR} \L_V \left(\frac{U_P}{\N}\right) \delta B_{ST}\cr
&& + \frac{\sqrt{-g}}{96} \eps^{MNPRST} H_{MNR} \L_U \left(\frac{V_P}{\N}\right) \delta B_{ST}
\eea
We now dualize $H$ in the two last lines, which then become
\bea
&& - \frac{\sqrt{-g}}{16} (H^{MNP} - 2 L^{MNP}) V^Q U_{MP} \delta B_{NQ}\cr
&& - \frac{\sqrt{-g}}{16} (H^{MNP} - 2 L^{MNP}) U^Q V_{MP} \delta B_{NQ}\cr
&=& \frac{\sqrt{-g}}{64} \eps^{NQMPRS} F_{RS} U_{MP} \delta B_{NQ} \cr
&& + \frac{\sqrt{-g}}{64} \eps^{NQMPRS} G_{RS} V_{MP} \delta B_{NQ}\cr
&& + \frac{\sqrt{-g}}{8} L^{MNP} U_{MP} V^Q \delta B_{NQ}\cr
&& + \frac{\sqrt{-g}}{8} L^{MNP} V_{MP} U^Q \delta B_{NQ}
\eea
The last two lines are identically zero as we show in appendix \ref{van}, so finally we get
\bea
\delta \L_{\I\I\I} &=& - \L_V \left(\frac{\sqrt{-g}}{48} \eps^{MNPRST}   H_{MNR} \frac{U_P}{\N}\right) \delta B_{ST}\cr
&& - \L_U \left(\frac{\sqrt{-g}}{48} \eps^{MNPRST} H_{MNR} \frac{V_P}{\N}\right) \delta B_{ST}\cr
&& + \frac{\sqrt{-g}}{32} \eps^{MNPRST} \delta B_{ST} \left(F_{MN} U_{PR} + G_{MN} V_{PR}\right)\cr
&& + \frac{\sqrt{-g}}{96} \eps^{MNPRST}   H_{MNR} \L_V \left(\frac{U_P}{\N}\right) \delta B_{ST}\cr
&& + \frac{\sqrt{-g}}{96} \eps^{MNPRST} H_{MNR} \L_U \left(\frac{V_P}{\N}\right) \delta B_{ST}
\eea

\subsection{Summing up the three variations} 
Let us sum up the three variations that we have got,
\bea
\delta \L_{\I} + \delta \L_{\I\I} + \delta \L_{\I\I\I} &=&  \Bigg[\L_V \left(\frac{\sqrt{-g}}{48} \eps^{MNPRST} H_{MNP} \frac{U_R}{\N} + \frac{ \sqrt{-g}}{8} \frac{H^{RST} U_R}{\N}\right) \cr
&& + \L_U \left(\frac{\sqrt{-g}}{48} \eps^{MNPRST} H_{MNP} \frac{V_R}{\N} + \frac{ \sqrt{-g}}{8}  \frac{H^{RST} V_R}{\N}\right)\cr
&& + \frac{\sqrt{-g}}{96} \eps^{MNPRST} \left(H_{MNR} \L_V \frac{U_P}{\N} + H_{MNR} \L_U \frac{V_P}{\N}\right)\cr
&&  - \frac{1}{8} \left(\L_V \sqrt{-g} - \Omega_V\sqrt{-g}\right) \frac{G^{ST}}{\N} - \frac{1}{8} \left(\L_U \sqrt{-g} - \Omega_U\sqrt{-g}\right) \frac{F^{ST}}{\N} \cr
&& - \frac{\sqrt{-g}}{4} P_Q^T \nabla_R H^{RSQ} \Bigg]\delta B_{ST}
\eea
We now see that in order for this variation to vanish we must impose the following geometric conditions,
\bea
\L_V \sqrt{-g} &=& \Omega_V\sqrt{-g}\cr
\L_U \sqrt{-g} &=& \Omega_U\sqrt{-g}\cr
\L_V \frac{U_P}{\N} &=& 0\cr
\L_U \frac{V_P}{\N} &=& 0
\eea
When these geometric conditions are satisfied, the variation reduces to 
\bea
\delta \L_{\I} + \delta \L_{\I\I} + \delta \L_{\I\I\I} &=&  \frac{1}{8} \Bigg[\L_V \left(\frac{\sqrt{-g}}{6} \eps^{MNPRST} H_{MNP} \frac{U_R}{\N} + \sqrt{-g} \frac{H^{RST} U_R}{\N}\right) \cr
&& + \L_U \left(\frac{\sqrt{-g}}{6} \eps^{MNPRST} H_{MNP} \frac{V_R}{\N} + \sqrt{-g}  \frac{H^{RST} V_R}{\N}\right)\cr
&& - 2 \sqrt{-g} P_Q^T \nabla_R H^{RSQ} \Bigg]\delta B_{ST}
\eea
It is also useful to rewrite this variation as
\bea
\delta \L &=& \frac{1}{4} \E^{ST} \delta B_{ST}
\eea
where
\bea
\mathcal{E}^{MN} &=& \mathcal{L}_V \left( \sqrt{-g} \frac{L^{MNP} U_P}{\mathcal{N}} \right) 
+ \mathcal{L}_U \left( \sqrt{-g} \frac{L^{MNP} V_P}{\mathcal{N}} \right) 
+ 2 \sqrt{-g} P^{[M}_Q \nabla_R L^{N]QR}
\label{eq:full_expr}
\eea
For any tensor $X^{MN}$ we have
\bea
\L_V (\sqrt{-g} X^{MN}) &=& \nabla_P (\sqrt{-g} V^P X^{MN}) + \sqrt{-g} (X^{NP} \nabla_P V^M - X^{MP} \nabla_P V^N)
\eea
By applying this formula, we have obtained after some computations the  result
\bea
\E^{MN} &=& \nabla_R \left(\sqrt{-g} \left(3Q^{[R}_P L^{MN]P} - 2 L^{MNR}\right)\right)\cr
&& + \sqrt{-g} L^{[N|PR} \left(U_{PR} V^{M]} + V_{PR} U^{M]}\right) 
\eea
but by the result in appendix \ref{van}, the second line vanishes identically. 

In conclusion, this rather long computation has led us to the following important result. Under a generic variation of the gauge potential, the Lagrangian varies as
\bea
\delta \L &=& \frac{1}{4} \E^{MN} \delta B_{MN}
\eea
where  
\bea
\E^{MN} &=& \partial_R \left(\sqrt{-g} T^{RMN}\right)
\eea
with
\bea
T^{RMN} &=& 3Q^{[R}_P L^{MN]P} - 2 L^{MNR}\label{T}
\eea
and specifically then, the Euler-Lagrange equations of motion are then given by 
\bea
\E^{MN} &=& 0
\eea

\section{The integrated equations of motion}
We may integrate the equations of motion locally and present them in the integrated form as
\bea
T^{MNP} &=& \frac{1}{2} \sqrt{-g} \eps^{MNPRST} \partial_R X_{ST}
\eea
where we will call $X_{MN}$ integration constants. These are not constants in the usual sense, but in a generalized sense where they are parameters that have to satisfy the constraint equation (\ref{konst}), which we will now proceed to show. 

We start with rewriting the integrated equation of motion as
\bea
2 \t{L^{MNP}} + 3 V^{[M} \t{L^{NP]}_U} + 3 U^{[M} \t{L^{NP]}_V} = \frac{1}{2} \eps^{MNPRST} \partial_R X_{ST}
\eea
where $L^{MN}_U = L^{MNP} U_P$ and the tilde means that we project on the transverse space. We show that this left-hand side equals $T^{MNP}$ in appendix \ref{trv}. In a form langauge this equation reads
\bea
T = 2 \t{L} + v \wedge L_U + u \wedge L_V &=& * dX
\eea
where $L_U := \t{\iota_U L} := \iota_U L - u \wedge \iota_V \iota_U L$ and similarly for $L_V$. Here $T$ and $L$ are viewed as three-forms and $L = (H-*H)/2$ is anti-selfdual off-shell. Let us now expand $L$ in transverse component fields,
\bea
L &=& \t{L} + v \wedge L_U + u \wedge L_V + u \wedge v \wedge L_{UV}
\eea
Here $\t{L},L_U,L_V$ and $L_{UV}$ are assumed to be transverse fields, so that their contractions with $U$ and $V$ vanish. The Hodge dual is 
\bea
* L &=& - \N u \wedge v \wedge *_4 \t{L} - \N v \wedge *_4 L_U + \N u \wedge *_4 L_V - \N *_4 L_{UV} 
\eea
where we used the rules
\bea
* \t{L} &=& - \N u \wedge v \wedge *_4 \t{L}\cr
* (u \wedge L_V) &=& \N u \wedge *_4 L_V\cr
* (v \wedge L_U) &=& - \N v \wedge *_4 L_U\cr
* (u \wedge v \wedge L_{UV}) &=& - \N *_4 L_{UV}
\eea
In order to satisfy $L = - *L$ we now see that we shall impose the following constraints off-shell
\bea
L_U &=& - \N *_4 L_U\cr
L_V &=& \N *_4 L_V
\eea
while $\t{L}$ and $L_{UV}$ are unconstrained. 

Now we will examine the integration constants. We begin by expanding them as
\bea
X &=& \t{X} + u \wedge X_V + v \wedge X_U + u \wedge v \wedge X_{UV}
\eea
Then we expand
\bea
d X &=& d_4 \t{X} + du \wedge X_V + dv \wedge X_U\cr
&& + u \wedge (\L_V \t{X} - d_4 X_V)\cr
&& + v \wedge (\L_U \t{X} - d_4 X_U)\cr
&& + u \wedge v \wedge (d_4 X_{UV} - \L_U X_V + L_V X_U)
\eea
Then we take the Hodge dual, 
\bea
* dX &=& - \N u \wedge v \wedge *_4 (d_4 \t{X} + u \wedge X_V + v \wedge X_U)\cr
&& + \N u \wedge *_4 (\L_V \t{X} - d_4 X_V)\cr
&& - \N v \wedge *_4 (\L_U \t{X} - d_4 X_U)\cr
&& - \N *_4 (d_4 X_{UV} - \L_U X_V + \L_V X_U)
\eea
We can now see that we shall constrain these integration parameters such that  
\bea
d_4 \t{X} + du \wedge X_V + dv \wedge X_U &=& 0\label{konst}
\eea
since $T$ on the left hand side of the equation of motion $T = * dX$ does not have any term that contains $u \wedge v$.

\section{A generalized gauge symmetry}
We will now show that the action has a rather large gauge symmetry under which $H$ transforms non-trivially. The usual situation is that in an abelian theory the gauge symmetry is $\delta B = \Lambda$ that is closed, $d\Lambda = 0$ so that the field strength is invariant, $\delta H  = 0$. Here we will show that the action has a generalized gauge symmetry $\delta B = \Lambda$ where $\Lambda$ itself is not closed but instead this condition is replaced by equation (\ref{klosed}) under which $H$ has a nonvanishing generalized gauge variation. 

We will then show that this generalized gauge symmetry variation, which does not have to be infinitesimal but can be arbirarily large because the theory is abelian, enables us to match the integration constants $X$ that can be arbitrarily large with these generalized gauge parameters. In this way we are able to view the integration constants as gauge parameters and see that in a particarly chosen gauge the integrated equations of motion become
\bea
T &=& 0
\eea
It is then a small step to show that this leads directly to the selfduality equation of motion for 
\bea
H &=& * H
\eea
It is in this sense with the help of a generalized gauge symmetry, that we can say that we derive the selfduality equation of motion from the action.  

Let us first expand a generic variation of the gauge potential in its transverse components,
\bea
\delta B &=& \Upsilon + v \wedge \alpha + u \wedge \beta + v \wedge u \gamma
\eea
that induces the variation
\bea
\delta H &=& d_4 \Upsilon + dv \wedge \alpha + du \wedge \beta\cr
&& - v \wedge (d_4 \alpha - \L_U \Upsilon)\cr
&& - u \wedge (d_4 \beta - \L_V \Upsilon)\cr
&& + v \wedge u \wedge (d_4 \gamma - \L_V \alpha + \L_U \beta)
\eea
of the field strength. Under this variation, the action varies as
\bea
\delta S = - \frac{1}{2} (T,\delta H) = - \frac{1}{2} (2\t{L} + u \wedge L_V + v \wedge L_U, \delta H)
\eea
We can then see that in order for the action to be invariant for a generic field $\t{L}$, the parameters shall be constrained so that 
\bea
d_4 \Upsilon + dv \wedge \alpha + du \wedge \beta &=& 0\label{klosed}
\eea
which is a generalized closed condition on the gauge parameters $\Upsilon$ in the transverse space. Further, in order for the action to be invariant for a generic off-shell selfdual field $L_V$ the gauge parameters have to be constrained such that $d_4 \beta - \L_V \Upsilon$ is anti-selfdual, while for a generic off-shell anti-selfdual field $L_U$ the gauge parameters have to be such that $d_4 \alpha - \L_U \Upsilon$ is selfdual. Finally there is no restriction on $d_4 \gamma - \L_V \alpha + \L_U \beta$ since $T$ does not contain a term $\sim u \wedge v$. 

We shall now obtain how the left-hand side in the equations of motion $T = *dX$ varies under this generalized gauge transformation. We begin with obtaining the gauge variation of the anti-selfdual projection of $H$,
\bea
\delta L &=& \frac{\N}{2} *_4 (d_4 \gamma - \L_V \alpha + \L_U \beta)\cr
&& - v \wedge \N*_4(d_4\alpha-\L_U\Upsilon)\cr
&& + u \wedge \N*_4(d_4\beta-\L_V\Upsilon)
\eea
From this, together with the general expansion
\bea
L &=& \t{L} + u \wedge L_V + v \wedge L_U + u \wedge v \wedge L_{UV}
\eea
we extract the gauge variations of the transverse components, 
\bea
\delta\t{L} &=& \frac{\N}{2} *_4 (d_4 \gamma - \L_V \alpha + \L_U \beta)\cr
\delta L_U &=& - \N*_4(d_4\alpha-\L_U\Upsilon)\cr
\delta L_V &=& \N*_4(d_4\beta-\L_V\Upsilon)
\eea
From these variations, we can assemble the gauge variation of $T$ 
\bea
\delta T &=& \N *_4 (d_4 \gamma - \L_V \alpha + \L_U \beta)\cr
&& - v \wedge \N*_4(d_4\alpha-\L_U\Upsilon)\cr
&& + u \wedge \N*_4(d_4\beta-\L_V\Upsilon)
\eea
We shall now match the integration constants with these gauge parameters that appear in the gauge variation of $T$. If we choose
\bea
X_{UV} &=& - \gamma\cr
X_V &=& \beta\cr
X_U &=& \alpha \cr
\t{X} &=& \Upsilon
\eea
then we can see that the gauge variation $\delta T$ becomes identical with $*dX$. This means that the integration constants can be absorbed into gauge parameters and then be eliminated by an inverse gauge variation. Thus for a suitable gauge choice, the rather complicated integrated equations of motion
\bea
T &=& * dX
\eea
can be reduced down to 
\bea
T &=& 0
\eea
From $T = 0$, we subsequently obtain the equation $L = 0$ which is nothing but the selfduality equation of motion $H = *H$. To see this, we only need to note that $
T = 2 \t{L} + u \wedge L_V + v \wedge L_U = 0$ implies that $\t{L} = L_V = L_U = 0$ separately and hence $L = \t{L} + u \wedge L_V + v \wedge L_U = 0$. 

Let us comment that for spacetime geometries with $du$ and $dv$ nonvanishing, we can not put the integration constants $\t{X}$ to be closed in the transverse space as one might have naively expected. However, since we match all four integration constants with gauge parameters that precisely matches the constraint equation (\ref{konst}) with the generalized closed condition in (\ref{klosed}), we do not need to first solve the constraint equation (\ref{konst}) for $\t{X}$ and then try to match the remaining independent integration constants with corresponding independent gauge parameters. Instead our matching is made implicitly.

\section{The $(1,0)$ supersymmetric extension}
There is a $(1,0)$ supersymmetric completion of the selfdual tensor field, comprised of one real scalar field $\phi$ and two symplectic-Majorana Weyl spinors $\lambda_I$ for $I = 1,2$ transforming as a doublet under an $SU(2)$ R symmetry. The on-shell supersymmetry variations are
\bea
\delta \phi &=& - i \bar\eps^I \lambda_I\cr
\delta B_{MN} &=& i \bar\eps^I \Gamma_{MN} \lambda_I\cr
\delta \lambda_I &=& \frac{1}{12} \Gamma^{MNP} \eps_I T_{MNP} + \Gamma^M \eps_I \partial_M \phi + 4 \eta_I \phi
\eea
where the two supersymmetry parameters are conformal Killing spinors
\bea
\nabla_M \eps_I &=& \Gamma_M \eta_I
\eea
Our gamma matrix conventions are summarized in appendix \ref{gammas}. Here $T^{MNP}$ is the tensor field that we obtained in equation (\ref{T}). The matter part of the action is given by
\bea
S_{matter} &=& \int d^6 x \L_{matter}
\eea
where the Lagrangian density is given by
\bea
\L_{matter} &=& - \frac{\sqrt{-g}}{2} g^{MN} \partial_M \phi \partial_N \phi - \frac{\sqrt{-g}}{10} R\phi^2 + \frac{i\sqrt{-g}}{2} \bar\lambda^I \Gamma^M \nabla_M \lambda_I
\eea
We find that its supersymmetry variation is given by 
\bea
\delta \L_{matter} &=& \frac{i\sqrt{-g}}{4} \bar\lambda^I \Gamma^{NP} \eps_I \nabla^M T_{MNP}\cr
&=& - \frac{i\sqrt{-g}}{4} \bar\eps^I \Gamma_{NP} \lambda_I \nabla_M T^{MNP}\cr
&=& - \frac{\sqrt{-g}}{4} \delta B_{NP} \nabla_M T^{MNP}
\eea
We also know that the variation of the tensor field part of the action has the general variation 
\bea
\delta \L_{tensor} &=& \frac{\sqrt{-g}}{4} \delta B_{NP} \nabla_M T^{MNP}
\eea
that we obtained previously through a lengthy computation. Of course since this is valid for an arbitrary variation of the gauge potential, it is in particular true for a supersymmetry variation of the gauge potential. Now if we add the supersymmetry variations together, we find that 
\bea
\delta \L = \delta \L_{tensor} + \delta \L_{matter} = 0
\eea
That is, the action is supersymmetric.

When we study the tensor field in the $2+4$ formulation we discover a set of geometric conditions:
\begin{enumerate}
\item The vector fields $U$ and $V$ must lightlike and commute.
\item Both $U$ and $V$ have to be conformal Killing vector fields.  
\item Their curvatures $dv$ and $du$ have to be selfdual and anti-selfdual respectively in transverse space. 
\end{enumerate}
We note that transversality of $dv$ and $du$, which means $\iota_V dv = \L_V v = 0$ and $ \iota_U dv = \L_U v = 0$, follows by just using the conformal Killing vector equation and the property that $U$ and $V$ commute, so this is not an independent requirement. Nevertheless, these are plenty of rather mysterious geometric conditions that we did not anticipate for a purely bosonic theory of a two-form gauge field. Although these geometric conditions do not mean that supersymmetry must exist, they get naturally explained if we do have supersymmetry. When we assume that the Lorentzian six-manifold has two symplectic-Majorana-Weyl conformal Killing spinors $\eps_I$ it was shown in \cite{Bak:2024ihe} that $V^M = \bar\eps^I \Gamma^M \eps_I$ and $U^M$ can be constructed from $\eta_I$ and that these vector fields satisfy all three properties enlisted above that are needed for the bosonic $2+4$ formulation of the selfdual tensor field.

\section{Discussion}
In \cite{Ko:2015zsy} an action was analyzed for the selfdual tensor field $H$ that splits the six-dimensional Minkowski spacetime $\mathbf{R}^{1,5}$ into $2+4$ dimentions so that the action has a reduced Lorentz symmetry only in these $2$ and $4$ dimensions separately. (A modified Lorentz symmetry can be obtained that rotates all six dimensions.) As an interesting result in \cite{Ko:2015zsy}, it was found that the two dimensions have to contain the time direction, while the four transverse dimensions shall be spatial dimensions. 

In this work, we have generalized \cite{Ko:2015zsy} to a curved spacetime. We have shown that the two dimensions shall be associated with two commuting lightlike conformal vector fields $U$ and $V$ that are such that $dv$ is selfdual and $du$ is antiselfdual in the transverse four-dimensional space. 

We find it surprising to get these geometric conditions on the six-manifold from the purely bosonic theory of a selfdual three-form when we apply the $2+4$ formulation. These conditions may explain why a curved spacetime generalization for the action could not be found in \cite{Ko:2015zsy} because we need to restrict the spacetime geometry somewhat. Why this restriction comes about here is a surprise in the bosonic theory because after all we can write down the Maxwell action for a non-chiral tensor field as $S = - \frac{1}{24} \int_M H \wedge *H$ on any Lorentzian six-manifold $M$ and of course we can also write down the selfduality equations of motion $H = *H$ on any Lorentzian six-manifold. So it is only when we ask for an action in the $2+4$ formulation that these geometric conditions emerge. Then do we need the $2+4$ formulation to exist, or can the $1+5$ formulation of Henneaux and Teitelboim be sufficient? In the work of Dirac \cite{Dirac:1949cp} there are three distinct categories for a Hamiltonian formulation of relativistic dynamics in Minkowski space, the instant form, the point form and the front form. Here the $1+5$ formulation corresponds to the instant form and the $2+4$ formulation corresponds to the front form and both formulations shall be equally valid and then we can not accept to not have the $2+4$ formulation in Minkowski spacetime. Once having said that, it seems quite natural to ask for the curved space generalization thereof to also exist.

\subsection*{Acknowledgements}
This work was supported in part by Basic Science Research Program through NRF funded by the Ministry of Education (2018R1A6A1A06024977).

\appendix

\section{Proof of an identity}\label{ide}
We prove here that we have the relations
\bea
H^{[M|PQ|} V^{N]} U_{PQ} &=& \frac{1}{2} \eps^{MNPQRS} F_{RS} U_{PQ}\cr
H^{[M|PQ|} U^{N]} V_{PQ} &=& \frac{1}{2} \eps^{MNPQRS} G_{RS} V_{PQ}
\eea
Let us begin with expanding 
\bea
\frac{1}{2} \eps^{MNPQRS} F_{RS} &=& \frac{1}{2} \eps^{MNPQRS} H_{RST} V^T \cr
&=& \frac{1}{12} \eps^{MNPQRS} \eps_{RSTUVW} H^{UVW} V^T + \eps^{MNPQRS} L_{RST} V^T\cr
&=& 2 H^{MN[P} V^{Q]} + 2 H^{PQ[M} V^{N]} + \eps^{MNPQRS} L_{RST} V^T
\eea
Then the desired identity follows upon contracting with $U_{PQ}$ and using $U_{PQ} V^Q = 0$,
\bea
H^{PQ[M} V^{N]} U_{PQ} + \frac{1}{2}\eps^{MNPQRS} L_{RST} V^T U_{PQ} &=& \frac{1}{2} \eps^{MNPQRS} F_{RS} U_{PQ}
\eea
up to the term
\bea
\eps^{MNPQRS} L_{RST} V^T U_{PQ}
\eea
that vanishes identically by assuming that $U_{PQ}$ is antiselfdual in the transverse four space, because $(L_V)_{RS}$ is selfdual there by construction.

\section{Vanishing expressions} \label{van}
Here we show that the following expressions are identically zero,
\bea
L^{M[N|P|} U_{MP} V^{Q]} &=& 0\cr
L^{M[N|P|} V_{MP} U^{Q]} &=& 0
\eea
Let us focus on the first identity since the second then follows by permuting $U$ and $V$. We start by dualizing the first expression thrice,
\bea
L^{M[N|P|} U_{MP} V^{Q]} &=& \frac{1}{288} \eps^{MEPRST} \eps_{ABCDEF} \eps^{ABCDNQ} L_{RST} U_{MP} V^{F}  \cr
&=& \frac{5}{12} \eps^{ABCDNQ} L_{[BCD} U_{FA]} V^F\cr
&=& \frac{1}{4} \eps^{ABCDNQ} L_{F[CD} U_{AB]} V^F\cr
&=& \frac{1}{4} \eps^{ABCDNQ} (L_V)_{[CD} U_{AB]} 
\eea
This vanishes by selfduality of $(L_V)_{CD}$ and antiselfduality of $U_{AB}$. For the other relation we permute $U$ and $V$ and then again we find that it vanishes by antiselfduality of $(L_U)_{CD}$ and selfduality of $V_{AB}$.

\section{Equivalence of two equations of motion}\label{trv}
Here we will show that the two expressions
\bea
T^{MNP}_{\I} &=& 2 L^{MNP} - 3 Q^{[M}_Q L^{NP]Q}
\eea
and 
\bea
T^{MNP}_{\I\I} &=& 2 \t{L^{MNP}} + 3 V^M \t{L^{NP}_U} + 3 U^M \t{L^{NP}_V}
\eea
are identical, $T^{MNP}_{\I} = T^{MNP}_{\I\I}$. Here ${L^{MN}_U} = L^{MNP} U_P/\N$ and ${L^{MN}_V} = L^{MNP} V_P/\N$ and the tilde means that we project these on the transverse space. All our expression are here implicitly understood to be antisymmetrized in $MNP$. 

We begin by substituting
\bea
Q^M_Q &=& \frac{1}{\N} (U^M V_Q + V^M U_Q)
\eea
in $T^{MNP}_{\I}$. Then we get
\bea
T^{MNP}_{\I} &=& 2 L^{MNP} - 3 (U^M L^{NP}_V + V^M L^{NP}_U)
\eea
Next we expand 
\bea
L^{MNP} &=& \t{L^{MNP}} + 3 V^M \t{L^{NP}_U} + 3 U^M \t{L^{NP}_V} + 6 U^M V^N \t{L^P_{UV}}
\eea
By contractions we obtain
\bea
\frac{U_M}{\N} L^{MNP} &=& \t{L^{NP}_U} - 2 U^{[N} \t{L^{P]}_{UV}}\cr
\frac{V_M}{\N} L^{MNP} &=& \t{L^{NP}_V} + 2 V^{[N} \t{L^{P]}_{UV}}\cr
\frac{U_M V_N}{\N^2} L^{MNP} &=& - \t{L^P_{UV}}
\eea
We now insert these expansions and we get
\bea
T^{MNP}_{\I} &=& 2 (\t{L^{MNP}} + 3 V^M \t{L^{NP}_U} + 3 U^M \t{L^{NP}_V} + 6 U^M V^N \t{L^P_{UV}})\cr
&& - \frac{3}{\N} (U^M L^{NP}_V + V^M L^{NP}_U)
\eea
Now we encounter a difference
\bea
12 U^M V^N \t{L^P_{UV}} - 6 U^M V^N \t{L^P_{UV}} + 6 V^M U^N \t{L^P_{UV}} &=& 0
\eea
Having canceled the components $\sim U^M V^N$ we are left with 
\bea
T^{MNP}_{\I} = 2 \t{L^{MNP}} + 3 V^M \t{L^{NP}_U} + 3 U^M \t{L^{NP}_V} = T^{MNP}_{\I\I}
\eea
as advertized.

\section{Our gamma matrix conventions}\label{gammas}
Our 6d gamma matrices will be denoted $\Gamma_M$ and the 6d charge conjugation matrix is denoted $C$. We have 
\bea
\Gamma_M^T &=& - C \Gamma_M C^{-1}\cr
C^T &=& C
\eea
These imply that $C \Gamma_M$ is an anti-symmetric matrix and $C \Gamma_{MN}$ is likewise antisymmetric. This means that for an anti-commuting supersymmetry parameter $\eps_I$ we have
\bea
\lambda^T_J C \Gamma_{MN} \eps_I &=& \eps_I^T C \Gamma_{MN} \eps_J
\eea
and consequently
\bea
\bar\lambda^I \Gamma_{MN} \eps_I &=& - \eps^I \Gamma_{MN} \eps_I
\eea

\end{document}